\documentclass[sigconf,nonacm,natbib=false]{acmart}
\makeatletter
\global\@ACM@balancefalse
\makeatother
\usepackage{balance}

\setcopyright{none}
\renewcommand\footnotetextcopyrightpermission[1]{}
\usepackage[
  backend=biber,
  style=numeric,
  sorting=none,
  giveninits=true,
  maxbibnames=3,
  minbibnames=3,
  doi=false,
  isbn=false,
  url=false,
  eprint=false
]{biblatex}
\DeclareNameAlias{author}{family-given}
\DeclareNameAlias{editor}{family-given}

\renewrobustcmd*{\bibinitperiod}{}
\renewrobustcmd*{\bibinitdelim}{}
\DeclareDelimFormat{multinamedelim}{\addcomma\space}
\DeclareDelimFormat{finalnamedelim}{\addcomma\space}
\DefineBibliographyStrings{english}{andothers={et\addabbrvspace al\adddot}}
\DeclareFieldFormat[article,inproceedings,incollection,thesis,misc]{title}{#1}
\DeclareFieldFormat{pages}{#1}

\AtBeginBibliography{\sloppy}

\newbibmacro*{compact:author-title}{%
  \printnames{author}%
  \setunit{\addperiod\space}%
  \printfield{title}}

\DeclareBibliographyDriver{article}{%
  \usebibmacro{bibindex}%
  \usebibmacro{begentry}%
  \usebibmacro{compact:author-title}%
  \newunit\newblock
  \printfield{journaltitle}%
  \setunit{\addcomma\space}%
  \printfield{year}%
  \iffieldundef{volume}
    {}
    {\setunit{\addcomma\space}%
     \printfield{volume}%
     \iffieldundef{number}{}{\printtext{\mkbibparens{\printfield{number}}}}}%
  \iffieldundef{pages}{}{\setunit{\addcolon\space}\printfield{pages}}%
  \usebibmacro{finentry}}

\DeclareBibliographyDriver{inproceedings}{%
  \usebibmacro{bibindex}%
  \usebibmacro{begentry}%
  \usebibmacro{compact:author-title}%
  \newunit\newblock
  \printfield{booktitle}%
  \setunit{\addcomma\space}%
  \printfield{year}%
  \iffieldundef{pages}{}{\setunit{\addcolon\space}\printfield{pages}}%
  \usebibmacro{finentry}}

\DeclareBibliographyDriver{thesis}{%
  \usebibmacro{bibindex}%
  \usebibmacro{begentry}%
  \usebibmacro{compact:author-title}%
  \newunit\newblock
  \printtext{Thesis}%
  \iflistundef{institution}{}{\setunit{\addcomma\space}\printlist{institution}}%
  \setunit{\addcomma\space}%
  \printfield{year}%
  \usebibmacro{finentry}}

\DeclareBibliographyDriver{misc}{%
  \usebibmacro{bibindex}%
  \usebibmacro{begentry}%
  \usebibmacro{compact:author-title}%
  \iffieldundef{howpublished}{}{\setunit{\addperiod\space}\printfield{howpublished}}%
  \iffieldundef{year}{}{\setunit{\addcomma\space}\printfield{year}}%
  \usebibmacro{finentry}}

\title{Research on Cross-media Science and Technology Information Data Retrieval}

\author{Yang Jiang}
\affiliation{%
  \institution{School of Computer Science (National Pilot School of Software Engineering), Beijing University of Posts and Telecommunications; Beijing Key Laboratory of Intelligent Telecommunication Software and Multimedia}
  \city{Beijing}
  \country{China}}

\author{Zhe Xue}
\authornote{Corresponding author.}
\affiliation{%
  \institution{School of Computer Science (National Pilot School of Software Engineering), Beijing University of Posts and Telecommunications; Beijing Key Laboratory of Intelligent Telecommunication Software and Multimedia}
  \city{Beijing}
  \country{China}}

\author{Ang Li}
\affiliation{%
  \institution{School of Computer Science (National Pilot School of Software Engineering), Beijing University of Posts and Telecommunications; Beijing Key Laboratory of Intelligent Telecommunication Software and Multimedia}
  \city{Beijing}
  \country{China}}

\begin{abstract}
Since the era of big data, the Internet has been flooded with all kinds of information. Browsing information through the Internet has become an integral part of people's daily life. Unlike news data and social data on the Internet, cross-media science and technology information data has different characteristics. This data has become an important basis for researchers and scholars to track current hot spots and explore future directions of technology development. As the volume of science and technology information data becomes richer, traditional science and technology information retrieval systems, which support only unimodal data retrieval and use outdated keyword-matching models, can no longer meet the daily retrieval needs of science and technology scholars. Therefore, in view of this research background, it is of profound practical significance to study cross-media science and technology information data retrieval systems based on deep semantic features, in line with domestic and international technology-development trends.
\end{abstract}

\keywords{technology information, cross media, semantic learning, retrieval and query}

\begin{document}
\maketitle

\section{Introduction}

Since the era of big data, increasingly rich data has flooded all aspects of life. The various data types presented on the Internet can be used to meet the needs of different users, and the Internet has entered people's lives and become an inseparable part of daily life. Unlike news and social information on the Internet~\cite{kou2016socialNetworkSearch}, cross-media science and technology information data has a different character. For scholar-centered scientific resources, multi-view scholar clustering with dynamic interest tracking can represent both multiple research perspectives and evolving interests~\cite{li2023scholarClustering}. This data contains large amounts of rich information, reflects real-time information hot spots, and includes considerable semantically similar information. However, because it is multi-source and multi-modal, designing a unified process for collecting, filtering, storing, and processing it, and then forming related business applications or products, is necessary to meet market demand.

As science and technology information data becomes increasingly abundant, traditional retrieval systems for scientific scholars~\cite{yang2015ontologyRetrieval} have gradually lagged behind because they support only keyword matching. They can no longer meet scholars' daily retrieval needs. Allowing researchers to retrieve more interesting and useful results from an ever-expanding volume places greater demands on retrieval systems. Interpretable machine-learning models can make the resulting intelligent decisions more transparent to users and system managers~\cite{li2019interpretableDecision}. At the same time, unimodal retrieval will gradually give way to inter-modal retrieval~\cite{liang2019deepQuantizationAttention}. Deeply integrating machine-learning and deep-learning algorithms with the characteristics of cross-media science and technology information data, and studying accurate cross-modal semantic-learning algorithms that support mutual retrieval, are therefore in line with domestic and international technology-development trends.

\section{Acquisition and Feature Extraction Analysis of Cross-media S\&T Information Data}

Technology-related resources continue to increase while stand-alone service systems have limited processing capacity. Distributed cluster technologies~\cite{xue2019deepLowRankMultiview,sun2009improvedKNN} are among the main implementations of current big-data technology~\cite{gu2018logBigDataSystem}. Cai~\cite{cai2017distributedETL} proposed an approach to industrial-data analysis using extract-transform-load (ETL) tools. ETL preprocesses data by merging multi-source data, analyzing it, reducing noise, and transforming dimensions. Hadoop MapReduce or Spark can be used for parallel processing to improve module performance~\cite{unknown2020parallelGeneticAlgorithm}. Other work~\cite{wen2018scholarUserPortrait,franceschet2017timeRank} describes converting scientific and technological data from different sources and structures into structured data, storing it in relational databases, and sampling key information. Deep modularity-based community detection offers a related mechanism for discovering coherent groups in heterogeneous information networks~\cite{yang2016modularityCommunity}. However, there is no mature crawling system for heterogeneous cross-media scientific and technological information data. Uniformly collecting, storing, and preprocessing such multi-source heterogeneous data remains urgent.

Vectorization is indispensable for feature extraction from cross-media technology information data~\cite{wu2022graphNeuralNetworksReview,guo2022distortedBanknote}. Semantic-similarity attention combined with hypergraph convolution can capture higher-order relations among scientific publications~\cite{li2026hypergraphPublication}. Machine-learning and deep-learning processing is needed to map the semantics of text and image resources~\cite{zhang2019technologyNewsTopicModel}. Teacher-student graph distillation provides a related way to recover incomplete features and structure before representation learning~\cite{huo2023t2gnn}. With the development of deep learning~\cite{ji2022deepLearningRobustness,bu2021attributeGraphClustering}, more abstract and higher-level representations have gradually replaced traditional machine-learning algorithms~\cite{liu2021deepMemoryNetworks,zhang2021imageInformationSentenceUnderstanding}. Consequently, text and image feature extraction based on deep learning has received growing attention.

Autoencoders can express distributed data compression accurately~\cite{eisa2017figurePlagiarism}. The TextCNN method applies convolutional neural networks to text feature extraction~\cite{zhao2016attentionCNNSentence}. Heterogeneous graph attention provides another option for semi-supervised classification when scientific descriptions are short and sparsely labeled~\cite{hu2019heterogeneousGAT}. Convolutional approaches~\cite{li2017varianceConstrainedState,li2017recursiveStateEstimation} do not perform as well for time-series text. Recurrent neural networks (RNNs) are therefore introduced for text feature extraction~\cite{zaremba2014rnnRegularization}; they perform well in sequence processing~\cite{zhao2017marketStateP2P} and more closely match human cognition. The memory and selective-forgetting concepts of long short-term memory and gated recurrent units~\cite{dey2017gruGateVariants,chung2017gatedRNNEvaluation} further improve on RNNs.

The BERT model~\cite{devlin2018bert} adopts the Transformer architecture~\cite{vaswani2017attention} to train language models. Retrieval-oriented masked-autoencoder pretraining can further adapt language representations to search tasks~\cite{xiao2022retroMAE}. BERT can add output layers for specific tasks without changing the pretrained model and can be fine-tuned for different applications. Long short-term memory handles and predicts important events by introducing memory cells~\cite{hochreiter1997lstm}. Bidirectional LSTM combines forward and backward LSTMs~\cite{jin2018bilstmChineseSegmentation}. ELMo uses a two-layer bidirectional LSTM to construct contextualized representations~\cite{sarzynskaWawer2021formalThoughtDisorder}, while BERT-Flow transforms anisotropic sentence-embedding distributions into smooth isotropic Gaussian distributions~\cite{li2020sentenceEmbeddings}.

For semantic representation of image resources, Simonyan and Zisserman proposed VGGNet in 2014~\cite{simonyan2014veryDeepCNN}. Convolutional networks are formed by stacking layers, enabling study of the relationship between network depth and performance. ZFNet uses visualization to reveal the functions of intermediate layers~\cite{zeiler2014visualizingCNN}. Dhankhar used ResNet-50 and VGG16 to recognize facial expressions and obtained good results on the KDEF dataset~\cite{dhankhar2019facialEmotionCNN}. Li and Xu proposed a dual-angle parallel-pruning optimization method~\cite{li2021vgg16DualAnglePruning}; pruning reduces VGG16's parameter count and improves feature-expression accuracy while maintaining training time. Omnidirectional image super-resolution with bi-projection fusion further illustrates how geometry-aware reconstruction can strengthen image representations~\cite{wang2024omnidirectionalSuperResolution}. Accurately extracting image and text features from science and technology information data thus remains a problem to be studied.

Current feature-extraction models for text and image resources have many parameters and deep network layers. They consume more server resources and take longer to infer. Filter-enhanced MLPs demonstrate that efficient models can encode ordered interaction signals without relying on deep recurrent stacks~\cite{zhou2022filterMLP}. Optimizing model structures and applying them in practical scenarios therefore remain important problems.

\section{Deep Semantic Learning of Cross-Media Science and Technology Information}

Semantic representation of cross-media scientific and technological information objects requires traditional machine-learning algorithms and deep-learning technologies to be integrated so that semantics from different modalities can be mapped to one another~\cite{yang2022semanticImageTextPatent,lu2022selfSupervisedSemanticHash,feng2021crossModalRetrievalReview}. When source data cannot be centralized, federated supervised cross-modal retrieval can learn aligned representations while limiting direct data exchange~\cite{li2024federatedCrossModal}.

Canonical correlation analysis (CCA) finds a subspace that maximizes pairwise correlations between two sets of heterogeneous data~\cite{hardoon2004canonicalCorrelation}. Joint feature selection and subspace learning jointly solve association-metric and coupled-feature-selection problems through an iterative algorithm~\cite{wang2015jointFeatureSubspace}. A correspondence autoencoder designs two single-modal encoder networks to construct a cross-media model~\cite{feng2014correspondenceAutoencoder}. Joint representation learning explores association and semantic information in a unified optimization framework~\cite{zhai2013jointCrossMediaRepresentation}. Deep semantic matching constructs a dual deep-neural mapping network to build a homogeneous semantic space~\cite{wei2016cnnCrossModalBaseline}.

Generative adversarial networks jointly construct a generator and discriminator to learn a distribution similar to a target~\cite{kurach2019ganRegularization,fang2020identityCycleGAN}. The generator analyzes sample-learning rules and produces samples through a neural model~\cite{xu2013imageFusionPCNN}; the discriminator determines whether generated data is real or synthetic. Continuous adversarial training changes the generator until it produces a distribution close to the target. Researchers have deeply integrated adversarial learning with cross-media semantic learning. A cross-media semantic-learning framework can use a feature mapper to confuse a modality classifier and form a modality-invariant representation~\cite{wang2016jointFeatureSubspace}. The modality classifier minimizes distances between similar semantic vectors across modalities through label predictions and triplet constraints, while interaction between the mapper and classifier maps different modalities into a common subspace.

Unsupervised cross-media retrieval through adversarial learning performs well with fewer annotations~\cite{he2017unsupervisedCrossModal}. Deep canonical correlation analysis effectively combines deep learning with correlation analysis~\cite{andrew2019deepCCA}. Optimized DCCA has also been used for cross-media semantic learning between text and image modalities~\cite{wang2018deepCCASemanticRetrieval}. CM-GANs simulate the joint distribution of different modalities with a generative network and form a generative model through weight-sharing constraints~\cite{peng2019cmGans}. CNN visual features provide another baseline for cross-media semantic learning~\cite{wei2017cnnCrossModalBaseline}; combining convolutional models with correlation learning can extract deep image features and improve retrieval~\cite{zou2018consistentRepresentationRetrieval}. Deep image features extracted by pretrained VGGNet can likewise be combined with homogeneous semantic algorithms~\cite{jin2018cnnCrossMediaApplication}.

Semantic-similarity-based adversarial cross-media retrieval constructs a semantic-similarity matrix in the feature-mapping network and learns through adversarial training~\cite{liu2021semanticSimilarityCrossMedia}. Semantics-adversarial and media-adversarial cross-media retrieval minimizes intra-media discrimination, inter-media consistency, and intra-semantic discrimination losses~\cite{li2022semanticsMediaAdversarial}. A prototype-based adaptive network uses a unified prototype to represent each semantic category across modalities, provides category-discriminative information, and adaptively learns cross-modal representations~\cite{zeng2021prototypeAdaptiveNetwork}. Federated graph neural networks further extend decentralized representation learning to cross-graph node classification~\cite{guan2021federatedGNN}. Reinforcement-based active client selection can improve participation decisions under heterogeneous graph distributions~\cite{wang2025reinforcementClientSelection}.

Research on cross-media scientific and technological information data is not yet mature. Proposing effective semantic-learning algorithms for this domain therefore remains an open problem.

\section{Cross-media Sci-tech Information Data Retrieval Based on Deep Semantic Features}

Search-engine technology dates back to around 1990, and image- and ciphertext-retrieval work illustrates the continuing breadth of retrieval tasks~\cite{wang2021fuzzyImageRetrieval,ji2021encryptedTopKRetrieval,he2019fasterRcnnhRetrieval}. Early search engines retrieved file names from FTP servers and returned the locations of matching files. After decades of development, search-engine technology continues to innovate, with the goal of providing users with better and more accurate results~\cite{duan2019elasticsearchMooc}. Generative recommendation models can further unify retrieval and ranking within a single generation process~\cite{zhang2025unifiedGenerativeRecommendation}.

As an open-source search engine, Lucene~\cite{sha2019luceneFullText} mainly contains index construction, search, and management modules~\cite{ding2016luceneRanking}. It creates dictionaries and indexes through syntactic analysis and language processing~\cite{zhu2018luceneInvertedIndex,liu2019luceneIndexFiles,hu2018anomalyKernelDensity}. Retrieval sets are then sorted by relevance, and Lucene's scoring mechanism supports comprehensive query services.

Search engines often deploy enormous amounts of data~\cite{liu2021elasticsearchTechnologyResources}, making efficient single-machine retrieval difficult. Distributed retrieval schemes therefore become necessary~\cite{zheng2010chineseInvertedIndex,dou2019distributedIndexQuery}. Elasticsearch and Solr rely on distributed indexing, load balancing, failover, and recovery to provide retrieval capabilities~\cite{tao2020distributedEcommerce}. For decentralized information networks, FedSIN uses federated self-adaptive learning to obtain representations without centralizing all data~\cite{li2026fedSIN}. Communication-efficient reinforcement federated learning can further reduce coordination costs through dynamic client selection and adaptive gradient compression~\cite{pan2025rfcsc}. Their retrieval process sorts correlations between a query and database items, returns items above a correlation threshold, and then updates the retrieval model using feedback.

Deep interest networks mine historical behavior, apply attention-based weighting, and support personalized retrieval~\cite{zhou2018deepInterestNetwork}. Self-supervised graph co-training can stabilize session-based representations through complementary graph views~\cite{xia2021graphCoTraining}. Deep learning has also been combined with text-retrieval systems~\cite{tang2019deepLearningTextRetrieval}, including distributed architectures based on Hadoop and Spark streaming~\cite{li2017distributedConsensusKalman}. Semantic retrieval can abstract the internal logic of scientific literature and use knowledge-unit similarity to improve information collection~\cite{li2019scientificLiteratureSemanticSpace}. Knowledge graphs can extract relationships among scientific and technological entities and support retrieval-system construction~\cite{ren2019medicalKnowledgeGraph}.

Current retrieval systems cannot intelligently search according to scholars' and users' interests. Integrating diverse interests with search terms is therefore an unresolved problem. In summary, cross-media scientific and technological information retrieval systems still face incomplete data collection, inaccurate semantic understanding, and limited ability to perform intelligent interest-aware retrieval.

\section{Conclusion}

The Internet is flooded with all kinds of information. In the era of big data, accessing information through the Internet has become an indispensable part of daily life. Cross-media science and technology information data differs from other Internet data because of its scientific and technological attributes. It has gradually become an important source for scholars exploring current technology hot spots and planning future research directions. As science and technology information becomes more abundant, traditional keyword-matching and unimodal retrieval methods have lagged behind and can hardly meet scholars' daily research needs. Against this background, this paper investigates semantic learning for cross-media science and technology information data.

\begin{acks}
This work is supported by the National Key R\&D Program of China (2018YFB1402600) and the National Natural Science Foundation of China (61772083, 61877006, 61802028, and 62002027).
\end{acks}

\balance
\printbibliography[title={References}]

\end{document}